\documentclass[journal=jacsat,manuscript=article,layout=twocolumn]{achemso}
\usepackage[version=3]{mhchem} 
\usepackage{amsmath}

\author{Ethan J. Taylor}
\affiliation[UAB]
{Department of Physics, University of Alabama at Birmingham, Birmingham AL}
\author{Vasudevan Iyer}
\affiliation[ORNL]
{Center for Nanophase Materials Sciences, Oak Ridge National Laboratory, Oak Ridge TN}
\author{Bibek S. Dhami}
\affiliation[UAB]
{Department of Physics, University of Alabama at Birmingham, Birmingham AL}
\author{Clay Klein}
\affiliation[ORNL]
{Department of Physics, Clarion University, Clarion PA}
\author{Benjamin J. Lawrie}
\affiliation[ORNL]
{Center for Nanophase Materials Sciences, Oak Ridge National Laboratory, Oak Ridge TN}
\affiliation[ORNL2]
{Materials Science and Technology Division, Oak Ridge National Laboratory, Oak Ridge TN}
\email{lawriebj@ornl.gov}
\author{Kannatassen Appavoo}
\affiliation[UAB]
{Department of Physics, University of Alabama at Birmingham, Birmingham AL}
\email{appavoo@uab.edu}
\phone{+001(205) 934-0340}

\title{Hyperspectral nanoscale mapping of hybrid perovskite photophysics at the single grain level}

\keywords{hybrid perovskite, cathodoluminescence, photon recycling}

\begin{document}







\begin{abstract}
  Hybrid organic-inorganic perovskites have drawn significant interest for applications in optoelectronics over the last few years. Despite rapid progress in understanding the photophysics of perovskite, there remains a need for improved understanding of the effect of microstructure on perovskite emission. Here, we combine unsupervised machine learning and cathodoluminescence microscopy of a prototypical hybrid perovskite film to decode photophysical processes that are otherwise lost with conventional Gaussian image processing.  Hyperspectral maps are decoded with non-negative matrix factorization, revealing components relating to primary band-edge emission, photon recycling, and defect emission. A blind-spectral non-negative matrix factorization procedure provides additional understanding of changes in an intermediate perovskite phase under electron beam exposure and illustrates how traditional Gaussian techniques may hide relevant emission features that are critical to the development of environmentally robust perovskite devices. 
\end{abstract}

\section{Introduction}
Over the last decade, hybrid organic-inorganic perovskites have garnered widespread interest for  applications in photovoltaics\cite{misra2015temperature,yuan2015photovoltaic,al2017hybrid,dualeh2014thermal,williams2016current,niu2015review}, sensing\cite{kakavelakis2018solution,xu2016ultrasensitive,yang2019facile}, and neuromorphic computing\cite{lin2015organic,xiao2016energy} because of their impressive electronic properties that include a high absorption coefficient, band-gap tunability, low exciton binding energy, and long carrier lifetimes and diffusion lengths. Despite the potential of hybrid perovskite devices, improved long-term stability and phase engineering will be essential to continue progress. Improved understanding and control of intra-grain structure has become critical to optimizing charge carrier dynamics to satisfy specific device requirements. Indeed, the exact role that grain boundaries play, beneficial\cite{Li2015_GB, Yun2015_GB} or detrimental\cite{Du2016_GB, Doherty2020_GB}, is yet to be fully understood.
\par
Various microscopy and spectroscopy techniques have been employed to characterize the effect of grain structure and morphology on charge carrier dynamics and transport. For example, transmission electron microscopy (TEM) has been used to characterize how lead-related defects are formed and how they are localizes at boundaries\cite{Alberti2019_PbI2}.  TEM has also been used to characterize a temperature-dependent phase transition from tetragonal \ce{MAPbI3} to trigonal \ce{PbI2}\cite{Fan2017_TEM}, with some recent work pointing towards humidity as a driving factor in how grain boundaries influence film degradation\cite{Wang2017_GB}. Furthermore, various emission spectroscopies have been used to map the excited-state energetics and dynamics in grain interiors and grain boundaries through intensity and lifetime measurements\cite{Mamun2017_PL,Grancini2015_PL,Adhyaksa2018_GB}. Other methods including atomic force microscopy, confocal microscopy, electron backscatter diffraction, and computational techniques have been used to better understand the effect of grain boundaries on the optoelectronic properties of perovskite thin films\cite{Adhyaksa2018_GB, Shao2016_AFM, Song2020_AFM, deQuilettes2015_CM, Long2016_GB,Sherkar2017_GB, Park2019_GB}. Cathodoluminescence (CL) microscopy now serves as a powerful tool for high resolution mapping of hybrid perovskite photophysics~\cite{Guthrey2020_CL}, but the electron beam is not a completely non-perturbative probe.  Large electron beam energies and currents can locally heat and decompose perovskite films, resulting in increased \ce{PbI2} luminescence and in the emergence of an unstable intermediate perovskite phase that emits light at higher energy than the primary perovskite band-edge emission~\cite{Xiao2015_CL,Guthrey2020_CL}.
\par
In this Letter, we employ CL microscopy and machine-learning techniques to provide insights into the photophysical properties of a single grain of a prototypical \ce{MAPbI3} perovskite with a spatial resolution of better than 10 nm. The raw CL spectrum images highlight the complex spectral variation within the grain interior and at its boundary. Secondary electrons are measured concurrently with the CL in order to provide correlated maps of the grain morphology and the perovskite photophysics. We also systematically vary the electron beam energy in order to probe both depth-dependent effects\cite{Bischak2015_CL} and beam induced perovskite decomposition\cite{Xiao2015_CL}. The raw CL spectrum images are analyzed using non-negative matrix factorization (NMF) techniques that allows us to map spectral components across the grain. This unsupervised analysis highlights how degradation, as probed by \ce{PbI2} formation, is more likely to occur near grain boundaries. Increasing the beam energy initially increases the amount of damage at the boundaries of the grain, though additional increases in beam energy and current result in damage to the center of the grain. Furthermore, we also show how primary band-edge emission is distributed within the body of the grain and how the morphology of the thin film dictates nanometer scale regions where photon recycling takes place. A secondary characterization technique using Gaussian-restricted NMF shows that, in contrast to recent reports\cite{Mamun2017_PL}, the ``disordered'' phase for these perovskites appears to be uncorrelated with the \ce{PbI2} grain boundary behaviour. 
\section{Cathodoluminescence Microscopy}
Cathodoluminescence microscopy is now a workhorse tool for the characterization of nanophotonic materials. When a high-energy converged electron beam interacts with a sample, the transmitted, secondary, and backscattered electrons are routinely used to characterize the sample morphology.  The CL signal can provide a clear nanoscale picture of the energetics and dynamics of excited states in materials; it complements electron-energy loss spectroscopies (EELS) because EELS measures the energy lost by the electron beam as it interacts with matter while CL characterizes the photons emitted by matter after electron excitation.  CL microscopy is often delineated into coherent and incoherent CL\cite{brenny2014quantifying}. Coherent CL, so named because the emitted photons have a defined phase relationship with the electron excitation, results from incident electrons polarizing material at material interfaces; it is frequently used to probe the local density of states of nanophotonic and plasmonic media\cite{schefold2019spatial,schilder2020phase,hachtel2019spatially,hachtel2018polarization}. Incoherent CL measures the spontaneous emission of excited states after the electron beam deposits energy in matter, and so is a powerful tool for measuring the nanoscale energetics and dynamics of defects and excitons in heterogeneous materials\cite{li2017phase,massasa2021thin,Guthrey2020_CL,feldman2018colossal,iyer2021near,Xiao2015_CL,bischak2017origin,Du2016_GB}.

\begin{figure*}
  \includegraphics[width=\textwidth]{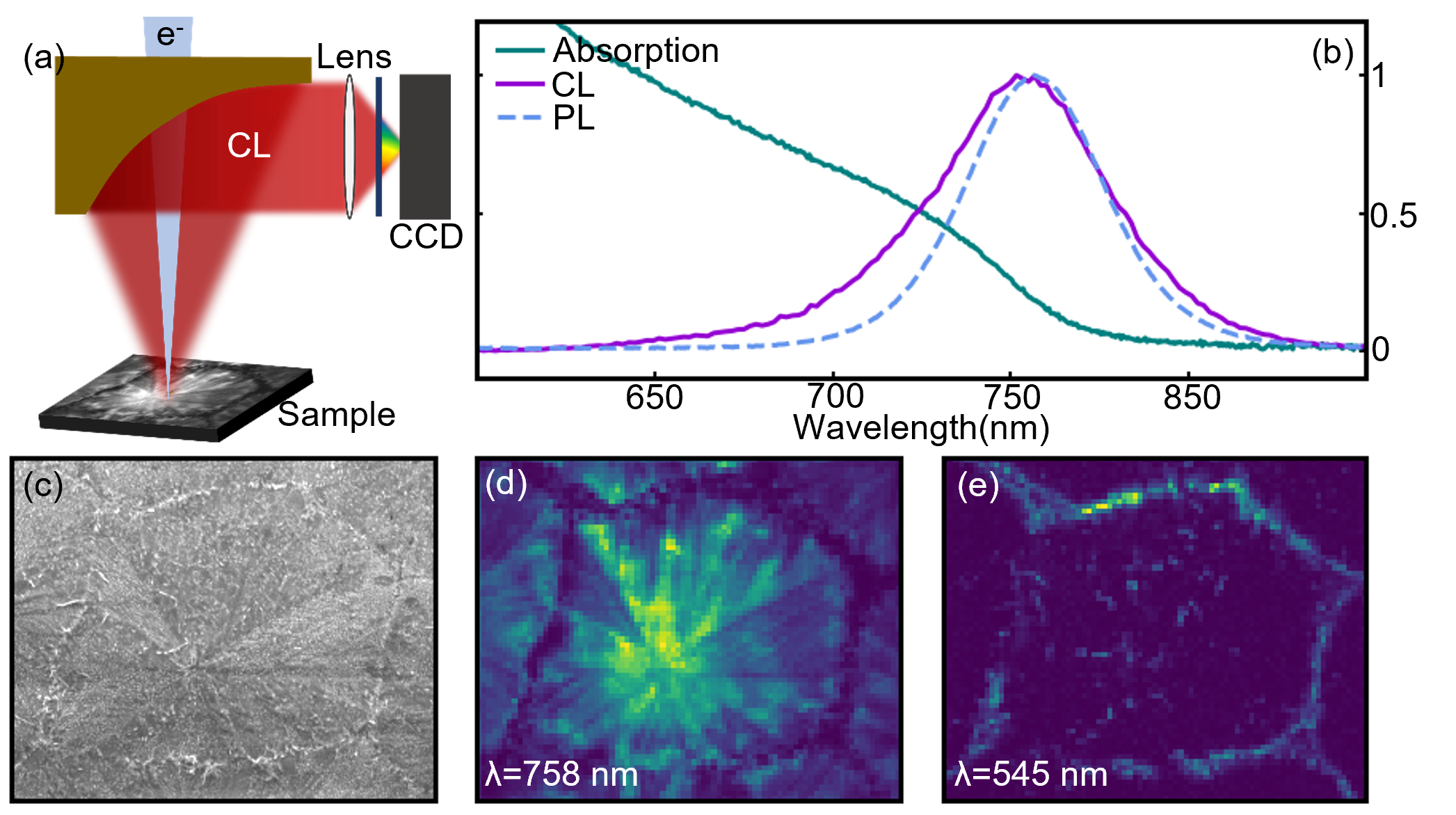}
  \caption{(a) Schematic of CL setup. The electron beam enters through a hole in a parabolic collection mirror before impacting the sample. The resulting photons are collected and directed through a focusing path before being spectrally dispersed onto a CCD. (b) Optical absorption, PL and CL at the center of a grain. (c) SEM images of sample collected with 5\,kV accelerating voltage with a horizontal field of view of 63\,$\mu$m. (d-e) CL intensity maps at 758\,nm and 545\,nm respectively.}
  \label{fig:schematic}
\end{figure*}

Here, as shown schematically in Figure \ref{fig:schematic}(a), we probe the incoherent CL generated by a perovskite thin film using a FEI Quattro environmental SEM with a Delmic Sparc CL collection module that utilizes a parabolic mirror to collect the CL generated by the film after electron beam excitation. An electron beam with beam energy of 5-20\,kV passes through a small hole in the parabolic mirror before exciting the sample with a spot size of order 1\,nm. The parabolic mirror then collects the resulting emission and directs it through a focusing lens into an Andor Kymera spectrograph. An Everhart-Thornley detector is used to measure the secondary electron (SE) signal concurrently with the CL signal. An SE map of a prototypical grain is shown in Figure \ref{fig:schematic}(c) for 5\,kV beam energy. Normalized optical absorption, photoluminescence(PL) and CL from the center of a grain are shown in Figure \ref{fig:schematic}(b) and CL maps at 758\,nm and 545\,nm are shown in Figure \ref{fig:schematic}(d) and (e) respectively. All CL measurements were performed with a beam current of 110\,pA. The 5 kV measurement utilized a 1\,s acquisition time per spectrum and the 10\,kV and 20\,kV measurements utilized 100\,ms acquisition times per spectrum. 

\section{Unmixing spectral components}

NMF is a powerful tool for spectral unmixing that separates overlapping signals in hyperspectral images. Here, we use an unsupervised ``blind-spectral'' NMF where the only physical constraint is that the spectral components must be non-negative\cite{Lee1999_NMF,Montcuquet2010_NMF} and a Gaussian-restricted NMF that adds additional constraints to the NMF decomposition based on initial Bayesian inference fitting methods.  All NMF decompositions were performed using the scikit-learn Python package.

\subsection{Gaussian NMF}
\begin{figure*}
    \centering
    \includegraphics[width=\textwidth]{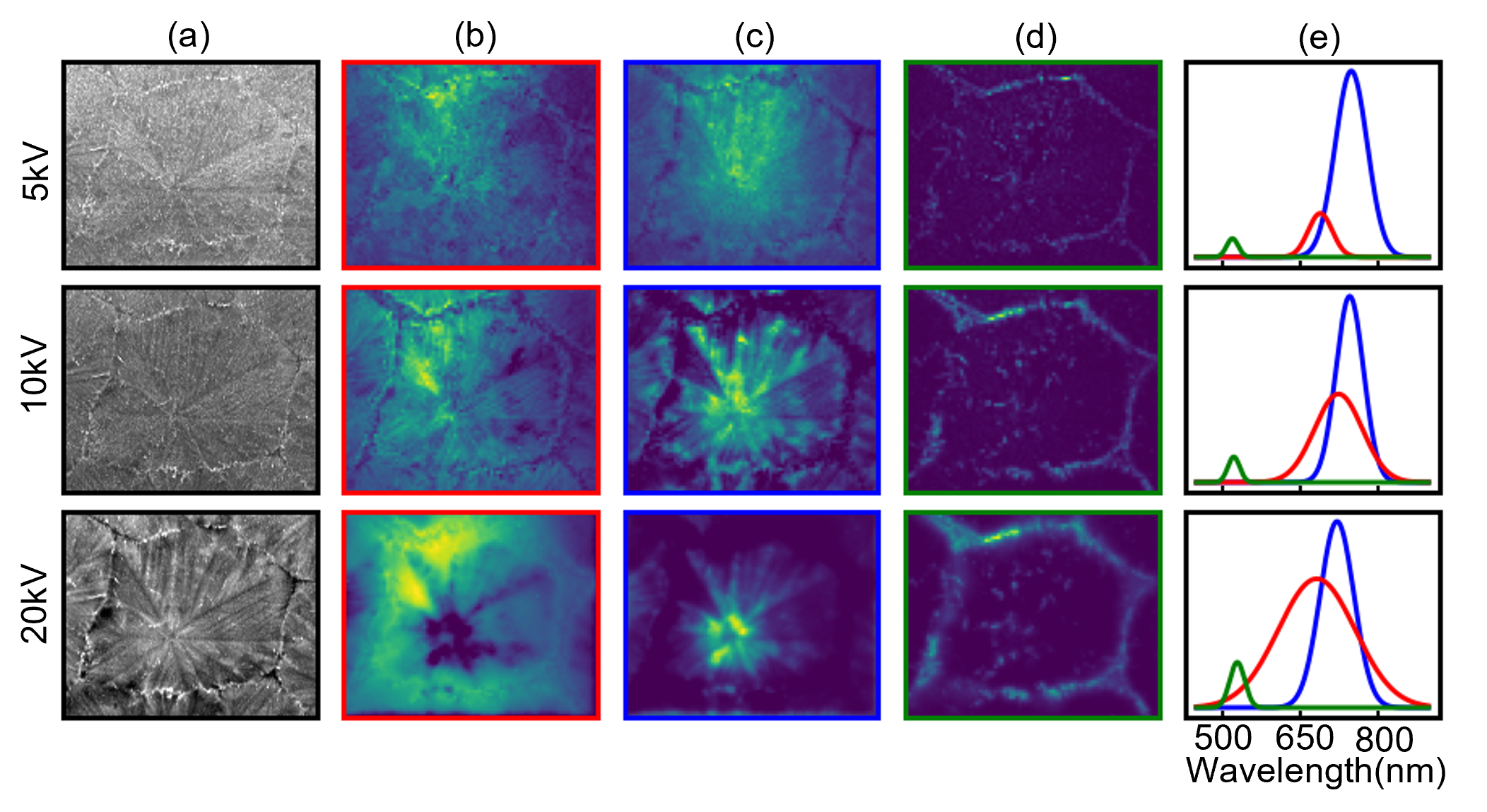}
    \caption{Gaussian restricted NMF for three different beam energies. (a) shows the corresponding SEM images collected simultaneously with CL. (b-d) Weight vectors with corresponding color for encoding matrix shown in (e). Captures (b) transitory damage states and primary band-edge emission, (c) primary band-edge emission and photon recycling, and (d) \ce{PbI2} emission suggestive of perovskite decomposition.}
    \label{fig:bayesian}
\end{figure*}

Previous reports of perovskite CL microscopies have relied on either maps of the CL intensity within given wavelength bins\cite{Xiao2015_CL} or least-squares Gaussian fits\cite{ummadisingu2021multi,jagadamma2021nanoscale}. The former can be a viable solution when the number of detected photons is too small to fit the CL spectra with high confidence. The latter can be computationally expensive, and it requires prior knowledge of the functional form of the CL spectrum, but it is sufficient for many applications. Here, we use a Bayesian inference fitting method\cite{pai2021magnetostriction} to fit three Gaussian modes to the average spectral response for each accelerating voltage and then perform a restricted NMF procedure using these fitted Gaussians as fixed row vectors in the encoding matrix while optimizing the weight matrix. The results of this Bayesian restricted NMF decomposition for a prototypical perovskite-grain CL spectrum image are shown in Figure \ref{fig:bayesian} for beam energies of 5, 10, and 20 kV. The images were acquired for the same grain beginning with the low energy electron beam and progressing to higher energy.

\subsection{Blind-Spectral NMF}
\begin{figure}
    \centering
    \includegraphics[width=\columnwidth]{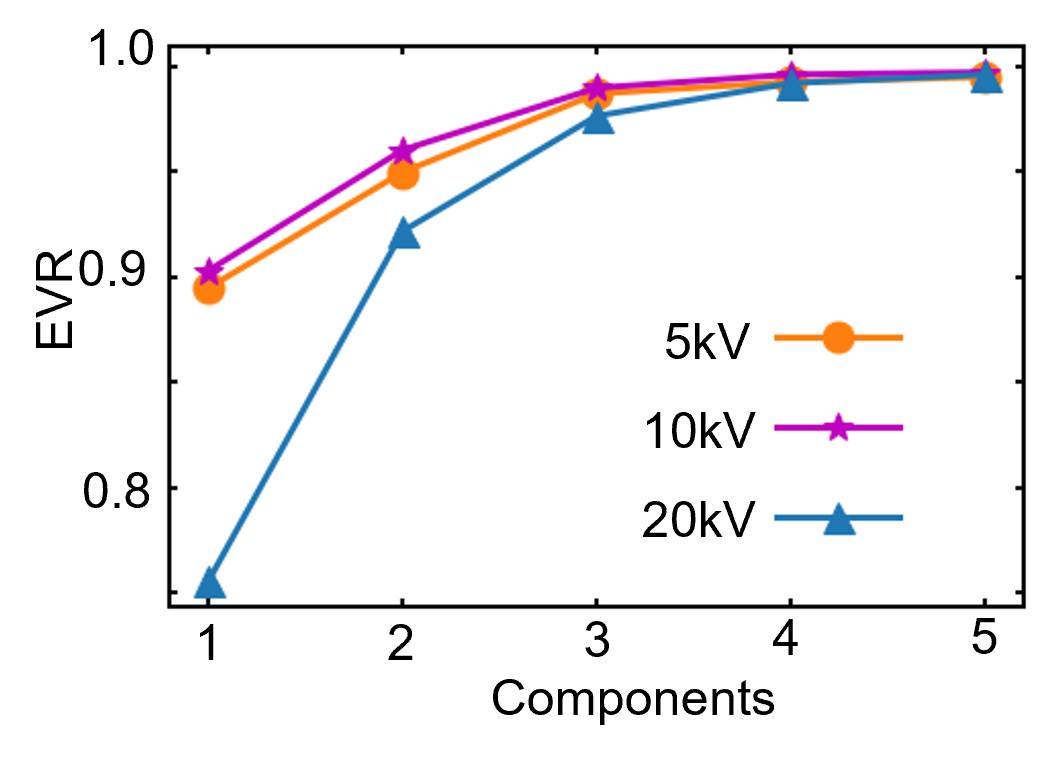}
    \caption{NMF explained variance ratio (EVR) for 1-5 components. Three components account for over 97\,\% of the variance.}
    \label{fig:variance}
\end{figure}

When performing blind-spectral NMF, we must first determine the appropriate number of components to include in the decomposition. Here, the explained variance ratio(EVR) was calculated for each accelerating voltage as shown in Figure \ref{fig:variance}. Because three components account for over 97\,\% of the variance in the data, no additional components were deemed necessary for the Blind-Spectral NMF calculations.
\par
\begin{figure*}
    \centering
    \includegraphics[width=\textwidth]{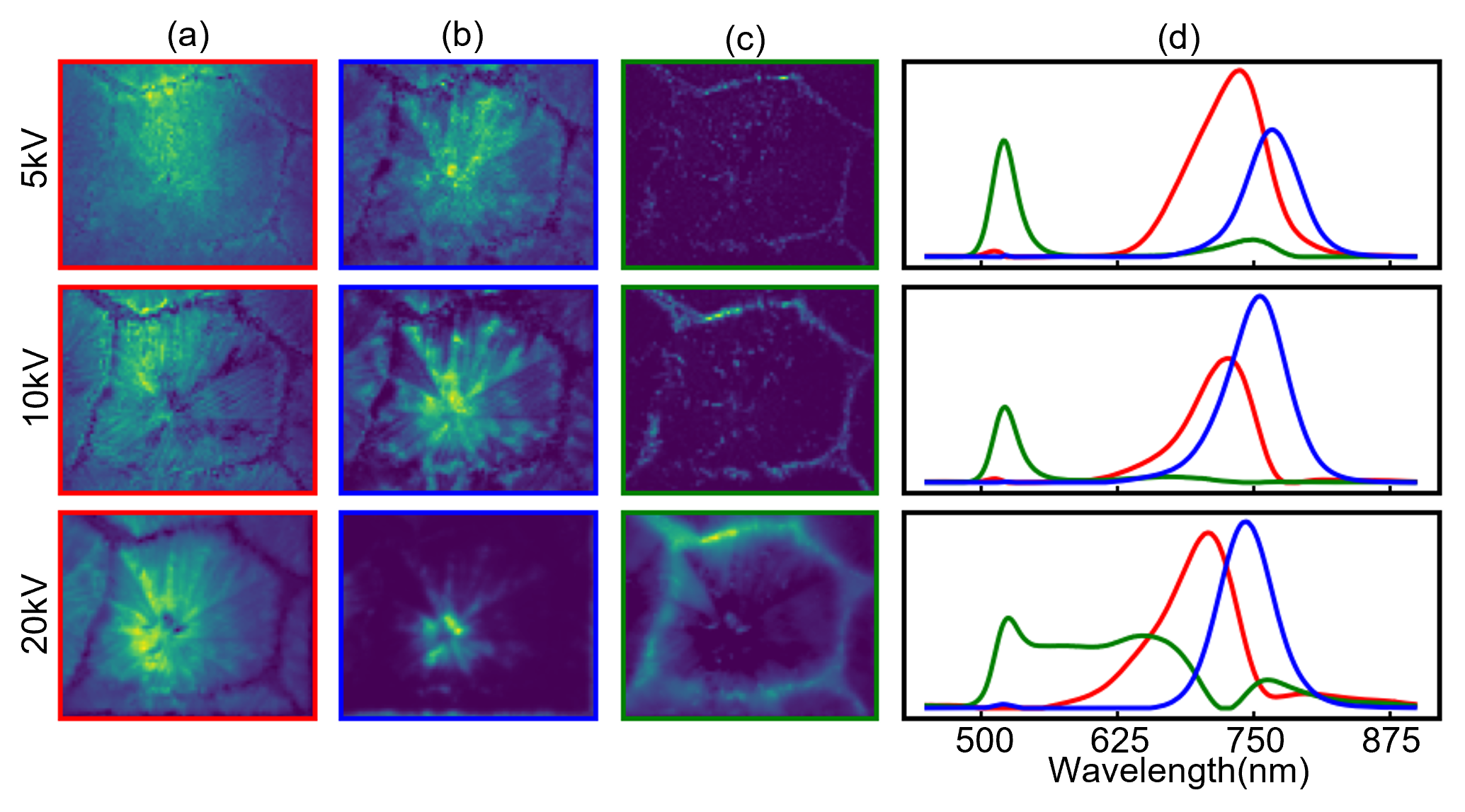}
    \caption{Blind-Spectral NMF of hybrid perovskite CL with varying beam energy. (a) NMF component corresponding to primary band-edge emission, (b) photon recycling, and (c) defect and intermediate phase emission. The normalized spectral components corresponding to the image decompositions are shown in (d). }
    \label{fig:NMF}
\end{figure*}

Figure \ref{fig:NMF} shows the results of the blind-spectral NMF for the same perovskite grain for the same three beam energies. As above, the spectral components shown in Figure \ref{fig:NMF}(d) correspond to the spatial maps in Figure \ref{fig:NMF}(a-c). Looking just at the low energy 5\,kV excitation, we see the NMF has decomposed the CL into three primary components, two localized in the grain interior and one at the grain boundary. As the accelerating voltage increases, corresponding to a greater depth of excitation\cite{Bischak2015_CL} and increased damage\cite{Xiao2015_CL}, the weights of these contributions vary. 
\par

\section{Discussion}
The SEM images in Figure \ref{fig:bayesian}(a) reveal a uniform grain structure for a grain of size $\sim$\,50\,$\mu$m, characteristic of \ce{MAPbI3} perovskite grains with a strong separation between grain interior and boundary. The grain structure is consistent with the proposed Voronoi growth detailed in past studies \cite{Hu2019_PerovskiteCrystal}, but as excitation depth increases, more intra-grain structures become visible. The CL spectrum images are well correlated with this intra-grain structure, so we argue that the crystallization process inside the grain strongly influences the emission parameters. However, because increased beam energy yields both increased electron-beam interaction volumes and increased damage, it is critical to minimize the electron-beam dose in order to facilitate depth dependent probes without causing excess damage. 

At 5\,kV, the electron beam probes the top 100\,nm of the film\cite{Guthrey2020_CL}, and we see that the primary band-edge emission contribution outweighs the photon recycling contribution in Figures \ref{fig:bayesian} and \ref{fig:NMF}. This aligns with understanding of the photon recycling process, as the amount of available material to contribute to the process is smaller and primary band-edge emission dominates across the surface of the grain in a uniform manner as shown in the spatial map. The photon recycling contribution is most prominent at the center of the grain and lacks any contribution at the grain boundary. The electron-beam conditions for the data reported here were intentionally chosen to minimize \ce{PbI2} damage, and the beam-induced damage is visibly minimized and localized to the grain boundary at 5\,kV. At 10\,kV, the weight of the two primary emission peak contributions flip, with the contribution assigned to photon recycling becoming the most prominent as a higher percentage of energy is deposited deeper within the grain. The contribution from primary band-edge emission localizes more along the interior of the grain in the spatial map and the photon recycling contribution remains strongest in the center of the grain. The \ce{PbI2} contribution grows slightly but remains localized at the boundary of the grain. At 20\,kV, the grain is now being damaged by the electron beam to a much larger degree. The primary band-edge contribution remains unchanged and is localized along the grain interior, with only slight variations due to the micro-structuring of the grain itself. Photon recycling can be seen to be entirely localized within the center of the grain. The intermediate degradation phase is substantially enhanced at 20 kV. This is consistent with past studies of the progression of damage in perovskite grains\cite{Guthrey2020_CL,Xiao2015_CL}, but it is captured here without distorting the contributions from relevant grain interior emission processes. 

Because the Gaussian-restricted NMF decomposition shown in Figure \ref{fig:bayesian} is constrained by Gaussian priors, it suppresses some spatial structure that is present in the raw hyperspectral CL maps. The same issue is present in the literature where frequentist Gaussian fits have been used to analyze CL spectrum images\cite{ummadisingu2021multi,jagadamma2021nanoscale}. The \ce{PbI2} CL (shown in Figure \ref{fig:bayesian}d) and the primary band-edge and photon recycling CL (shown in Figure \ref{fig:bayesian}c) remain well spatially resolved, but the CL from the intermediate perovskite phase (shown in Figure \ref{fig:bayesian}b) is poorly resolved because of the inappropriate choice of priors.

The blind spectral NMF shown in Figure \ref{fig:NMF} exhibits substantial similarities with the Gaussian-restricted NMF. However, the absence of Gaussian constraints allows for more complex spectral components to emerge in Figure \ref{fig:NMF}(d) and for spatial structure that is present in the raw CL hyperspectral data but suppressed by the Gaussian-NMF to emerge in Figure \ref{fig:NMF}(a-c). While the Gaussian constraints convolved the primary band-edge and photon recycling CL into a single Gaussian component, the blind spectral NMF shown in Figure \ref{fig:NMF} clearly delineates the primary band-edge emission in (a) from the photon recycling CL in (b). Moreover, the blind spectral NMF combines the \ce{PbI2} defect CL and the intermediate phase CL into a single non-Gaussian spectral component illustrated in Figure \ref{fig:NMF}(c). Because the three components used here yield a greater than 97\% EVR, it is reasonable to conclude that the \ce{PbI2} defect CL and the intermediate phase CL are in fact spatially correlated with one another because of an increased likelihood of defect formation at grain boundaries.  This is consistent with the limited literature that has explored beam-induced damage in hybrid perovskites~\cite{Xiao2015_CL,Guthrey2020_CL}.

\section{Conclusion}
In this study, we investigated the electron-beam energy-dependent hyperspectral CL of an \ce{MAPbI3} perovskite grain with a spatial resolution of 10nm using a combination of Gaussian-constrained NMF and blind-spectral NMF. We highlighted four prominent components of the spectrum images, including the primary band-edge perovskite CL, photon recycling CL, an intermediate phase associated with localized beam induced heating, and a fully decomposed \ce{PbI2} CL. While CL microscopy has become a well-used tool for probing perovskite films, the use of blind-spectral NMF allowed us to learn more about nanoscale perovskite photophysics, and, critically, to demonstrate a clear correlation between the intermediate perovskite phase and decomposed \ce{PbI2} CL near grain boundaries.  Indeed, this method allowed us to map the growth of the intermediate phase as a function of damage from the electron beam. This highlights the importance of understanding the optical properties of perovskite films near grain boundaries. Because the intermediate perovskite phase is chemically unstable, increased focus on understanding and minimizing the emergence of this intermediate phase will be critical to the development of environmentally robust perovskite devices.

\begin{acknowledgement}

Support for this project was provided by NASA EPSCoR RID (award number 80NSSC19M0051) and UAB startup funds. Cathodoluminescence microscopy was conducted at the Center for Nanophase Materials Sciences, which is a DOE Office of Science User Facility. BSD acknowledges financial support from the Alabama Graduate Research Scholars Program (GRSP) funded through the Alabama Commission for Higher Education and administered by the Alabama EPSCoR. We thank Dr. Sergei Kalinin for insightful discussion about data analysis.

\end{acknowledgement}




\bibliography{bibliography}

\end{document}